\pdfoutput=1
\documentclass[11pt]{article}
\usepackage[utf8]{inputenc}
\usepackage[margin=1in]{geometry}
\usepackage{amsmath, amssymb, amsthm}
\usepackage{physics}
\usepackage{graphicx}
\usepackage{booktabs}
\usepackage{enumitem}
\usepackage{bm}
\usepackage{float} 
\usepackage{booktabs,array}

\title{Centipede’s Leap into the Quantum Realm}
\author{
Kaytki Chakankar$^{1}$, Xinhui Tang$^{2}$, and Yiguo Zhang$^{3}$\\[4pt]
\small $^{1}$Notre Dame High School, San Jose, \texttt{kaytki.chak@gmail.com}\\
\small $^{2}$Cate School, \texttt{huihuixct@gmail.com}\\
\small $^{3}$Woodside Priory School, \texttt{yiguoseondpersonal@gmail.com}
}
\date{}

\usepackage[hidelinks]{hyperref}

\begin{document}
\maketitle

\begin{abstract}
\noindent The centipede game is a two-player non-zero-sum game. Each turn, a player can choose whether they want to take or pass a growing reward. The classical, rational solution of this game shows defection in the first round, when in reality, players cooperate much more often. Inspired by prior work employing quantum strategies in the prisoner’s dilemma, we showed that when similar quantum mechanics principles are applied to the centipede game, it leads to two new quantum Nash equilibria that are superior to the classical solution. Furthermore, by implementing our algorithm on Qiskit, we confirmed that leveraging quantum strategies, rather than strategies like backward induction, to solve the centipede game provided better payoffs for both players and more accurately modeled the game's real-life outcomes. Ultimately, we propose a generalized conjecture for similarly structured quantum games.
\end{abstract}

\section{INTRODUCTION}
Game theory has a wide range of applications in various fields such as economics, law, evolutionary biology, and computer science. In this paper, we’ll focus on the algorithmic side of game theory, particularly in its application to analyzing decisions and strategies. The two main categories of games in which game theory is applied are cooperative and non-cooperative games. Our emphasis will be on non-cooperative games, where each player chooses their strategy independently of others. An essential component in these games is that each player assumes the other player(s) will act rationally.

The first official research paper, \textit{Zur Theorie der Gesellschaftsspiele}, about classical game theory, was published by John von Neumann in 1928 [1]. A key idea that von Neumann introduced was the Minimax theorem. The Minimax theorem proves that in every two-player zero-sum game, an equilibrium can be reached if both players use mixed strategies. The minimax theorem inspired George Dantzig to coin the term linear programming in 1947 [2]. Dantzig then created the simplex method, which was a prevalent algorithm to solve linear programming problems, especially concerning optimization [3]. The simplex method was used for a variety of optimization problems, most notably in the US Air Force during WW2 [4]. In 1964, Carlton Lemke and Joseph Howson built on the pivoting structure of the simplex method. However, instead of using it for optimization problems, Lemke and Howson used it for finding the Nash equilibrium in a non-zero-sum two-player game [5]. Nash equilibrium is the idea that changing one’s strategy won’t result in a better personal outcome [6]. It is considered to be the most stable outcome of a strategic interaction. These developments laid the foundation for classical game theory.

Recent decades have seen the emergence of quantum mechanics as a powerful tool for re-imagining these models. Quantum mechanics emerged in the early 20th century to explain phenomena that classical physics couldn't, such as blackbody radiation and the photoelectric effect [7]. Many key findings made by scientists like Schrödinger, Einstein, and Heisenberg led to a new mathematical framework for describing particles at the sub-atomic level, which was necessary for advancements in quantum computing. The original idea that quantum computers could be superior to their classical counterparts came from Richard P. Feynman [8], who argued that quantum systems couldn’t be effectively represented by classical computers. He introduced the concept of quantum computers as a necessary tool for modeling nature. Scientists were eventually able to build on his work and produce core ideas, such as superposition and entanglement. These ideas are now being applied to rethink decision-making processes in fields like economics, cryptography, and game theory.

Quantum game theory is an extension of classical game theory, incorporating concepts from quantum mechanics (entanglement, unitary operations, tensor products, circuits), to rethink strategic decisions in both zero-sum and non-zero-sum games [9]. Instead of using discrete or probabilistic choices, players use qubits to evaluate multiple strategies at once with superposition (modeled with qubits being in two states at once) and apply unitary operations (representations of players’ decisions) as strategies. Entanglement can be used to model how one player’s choice can instantly affect the strategy of others, even without communication, and tensor products can be used to model the full state of a game, with all players’ strategies represented.

In the past, many studies have been done researching quantum strategies in all types of games. One of the first papers on quantum strategy was written by David A Meyer. Meyer mainly explored the game theory of the quantum penny flip game, a zero-sum game [10]. He states, “We prove that in general a quantum strategy is always at least as good as a classical one”. Following Meyer’s work, the field of quantum game theory rapidly expanded, showing that the possible strategies players can implement drastically increase when applying the principles of quantum mechanics. Some papers [11, 12] show that new Nash equilibria can be achieved when using quantum strategies and outperform classical models in certain situations.

This shift is relevant in games like the centipede game, a non-zero-sum game where two players take turns deciding whether to pass or take an increasingly large pot of money. The game ends once a player decides to defect and take the money, or when both players cooperate until the end of the game (the number of rounds may vary). Backward induction is an example of a possible strategy to find the Nash equilibrium for the centipede game. Although there are many solutions to find the proposed Nash equilibria for the centipede game, backward induction is often the most famous and referenced one. Backward induction is a method where you start reasoning from the end of the game to the start of the game to determine their best course of action [13]. Backward induction relies on the fact that players will assume their opponents will always act in their best interest, therefore making the future of the game predictable. In the centipede game, a player will always choose to defect and take the money in the first round by using backward induction.

\begin{figure}[H]
    \centering
    \includegraphics[width=0.5\linewidth]{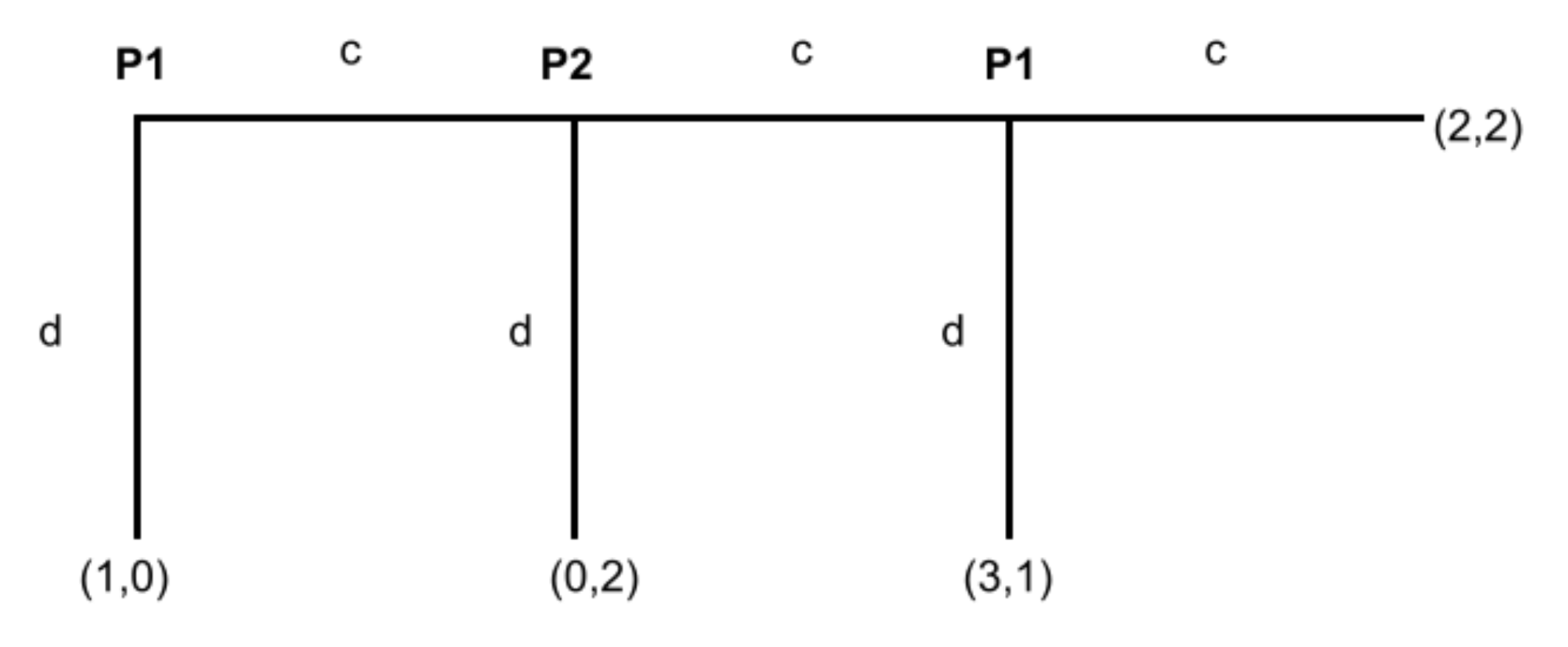}
    \caption{Illustration of the 3 rounds played in the centipede game}
    \label{fig:centipede3}
\end{figure}

Backward induction for the centipede game looks like this: We start at the last round. Since P1 can obtain a payoff of 3 by defecting, P1 would choose to defect in round 3. Then, P2 knows that rationally, P1 would defect at round 3, which would result in P2's payoff being 1; thus, P2 would defect at round 2 so that its payoff is 2. Consequently, P1 would predict P2 would defect at round 2, giving P1 a payoff of 0; therefore, P1 will always defect in the first round to get a payoff of 1.

Studies show that the centipede game is known for the gaps between its theoretical predictions and its experimental results, as players in real life do not follow the strategy of backward induction [14, 15]. However, we believe that leveraging quantum strategies that have been used to solve other classical games will give us an alternative Nash equilibrium that yields better outcomes for both players and will model how the game will play out in real life.

Quantum strategies have been proven to outperform classical alternatives in certain circumstances [16, 17, 18]. They lead to better joint outcomes due to their initial entangled state. The mechanism of entanglement (having the measurement of one qubit to instantly affect the state of another) allows for different actions and events to affect each other in a way that is classically unreproducible, leading to more cooperation in different games. An example of this is in the quantum version of the Prisoner’s Dilemma [19]. The Prisoner’s Dilemma is a game in which two players are presented with the option to rat out their partner (defect) or remain silent (cooperate) during an interrogation. If both players defect, they will have minimal punishment. However, if one player chooses to defect and one cooperates, the one who defects doesn’t receive a punishment, while the player who cooperates gets a considerable punishment. A study on the quantum algorithm used for the Prisoner’s Dilemma explains the algorithm that was used to achieve the best outcome in the Prisoner’s Dilemma. The study uses the Eisert-Wilkins-Lewenstein (EWL) Protocol, a well-known framework for quantizing games with 2 players and one round. The best result that the study got was if both players cooperated as explained above.

The EWL Protocol inspires a natural extension to more complex games. In particular, while classical backward induction in the centipede game only benefits one player, we hypothesize that applying quantum strategies to the centipede game will result in a more cooperative and rewarding outcome for both players. The quantum strategy would align closer with how the game might unfold in real life.

There has been previous work applying quantum mechanics to the centipede game. For their centipede game [20], Frackiewicz et. al. adapted the Marinatto–Weber scheme (a scheme in which a player can choose to play a classical game or its quantum counterpart) and found pure-strategy equilibrium to be (4.5,4.5), but this is purely theoretical. Additionally, the equilibrium is not “play-on to the very end”; Each individual run collapses to either (6, 4) or (3, 5); averaging the two classical payoffs with equal probability gives the average of (4.5, 4.5). We wish to address these drawbacks in this paper.

In this paper, we explore how a quantum circuit-based algorithm approaches the Centipede Game differently from classical methods, particularly in terms of outcomes. By modifying the EWL protocol from the prisoner’s dilemma [19], we find the mathematical solutions to the centipede game. We then model multiple rounds of the Centipede Game using Qiskit, a Python-based quantum computing framework, to show the feasibility of such a quantum approach in solving the Centipede Game.
%
%
%
%
%
%
%
\section{METHODOLOGY}
\subsection*{Math}
We adapted the Eisert--Wilkens--Lewnstein (EWL) protocol~[19] to the centipede game. The EWL protocol was originally developed to fit a 2-player, 1-round prisoner’s dilemma game, so we modified its framework notably to fit a 2-player, multi-round game like the centipede game; we call it the CTZ protocol.

To begin, each of the three rounds is represented by a qubit. Each qubit has to be entangled for the outcomes of each round to be able to affect each other. Quantum entanglement, a key factor that is used in quantum game theory, changes how each player acts. To entangle the decisions together, each qubit has to be in a state of superposition. Mathematically, the \(J\) entangler would allow us to turn unentangled qubits \(\ket{000}\) into
\[
\ket{\psi_{\text{unentangled}}}=\ket{000},\qquad
\ket{\psi_{\text{entangled}}}
= \frac{\ket{000}+ i\ket{111}}{\sqrt{2}}
= \frac{1}{\sqrt{2}}\ket{000} + \frac{1}{\sqrt{2}}\,i\ket{111}.
\]

In circuits, the \(J\) entangler can be replicated by applying a Hadamard gate on the first qubit, then two CNOT gates on 1st\(\to\)2nd qubit and 2nd\(\to\)3rd qubit, and finally an \(S\) relative phase gate.

\(\hat U\) is a unitary operator where the player’s decision could be encoded as a superposition of both cooperate and defect. In practice, we can set any of the two quantum gates to perform the operation, such as \(R_x(\Theta)\), \(R_y(\Theta)\), producing similar results, but we chose to use \(R_y(-\Theta)=\hat U\) in reference to EWL.
\[
\hat U(\theta,\phi)
=
\begin{bmatrix}
e^{i\phi}\cos(\theta/2) & \sin(\theta/2)\\[2pt]
-\sin(\theta/2) & e^{-i\phi}\cos(\theta/2)
\end{bmatrix},
\qquad
\hat U\!\left(\theta,\frac{\pi}{2}\right)
=
\begin{bmatrix}
i\cos(\theta/2) & \sin(\theta/2)\\[2pt]
-\sin(\theta/2) & -i\cos(\theta/2)
\end{bmatrix}.
\]
This is the quantum gate mentioned by EWL. However, since we entangled our game with a relative phase (the \(i\) in front of \(\ket{111}\)), the imaginary number in the matrix is redundant and is equivalent to
\[
\hat U(\theta,0)
=
\begin{bmatrix}
\cos(\theta/2) & \sin(\theta/2)\\[2pt]
-\sin(\theta/2) & \cos(\theta/2)
\end{bmatrix},
\]
which is much easier to use, since it's all in the reals, and this would still represent maximum entanglement.

After applying \(\hat U\) to the states of cooperation (\(\ket{0}\)) and defection (\(\ket{1}\)), we take the tensor product of all three rounds. Performing the tensor product is necessary to combine the three qubits, which represent the three rounds in the centipede game. To take the tensor product, we first have to find the probability each decision has to show cooperation or defection, as shown below:
\[
\alpha=\cos(\theta/2),\qquad \beta=\sin(\theta/2),
\]
\[
\begin{aligned}
U_1\ket{0}&=\alpha_1\ket{0}-\beta_1\ket{1}, &\quad U_2\ket{0}&=\alpha_2\ket{0}-\beta_2\ket{1}, &\quad U_3\ket{0}&=\alpha_3\ket{0}-\beta_3\ket{1},\\
U_1\ket{1}&=\beta_1\ket{0}-\alpha_1\ket{1}, &\quad U_2\ket{1}&=\beta_2\ket{0}-\alpha_2\ket{1}, &\quad U_3\ket{1}&=\beta_3\ket{0}-\alpha_3\ket{1}.
\end{aligned}
\]
The subscripts show the round number. Each round starts in the default state of cooperation. The alphas and betas represent the probability of each round ending in either cooperation or defection. The subscript on alpha and betas represents the respective \(\theta\) variable.

We take the tensor product of the unitary operators to our entangled qubits, which is displayed below:
\[
\ket{\psi_{\text{entangled}}}
= \frac{1}{\sqrt{2}}\bigl(U_1\ket{0}\otimes U_2\ket{0}\otimes U_3\ket{0}\bigr)
+ \frac{i}{\sqrt{2}}\bigl(U_1\ket{1}\otimes U_2\ket{1}\otimes U_3\ket{1}\bigr).
\]

\paragraph{1) Expansion of \(U_1\ket{0}\otimes U_2\ket{0}\otimes U_3\ket{0}\).}
\[
\begin{aligned}
U_1\ket{0}\otimes U_2\ket{0}\otimes U_3\ket{0}
&=(\alpha_1\ket{0}-\beta_1\ket{1})\otimes(\alpha_2\ket{0}-\beta_2\ket{1})\otimes(\alpha_3\ket{0}-\beta_3\ket{1})\\
&=\alpha_1\alpha_2\alpha_3\ket{000}-\alpha_1\alpha_2\beta_3\ket{001}-\alpha_1\beta_2\alpha_3\ket{010}
+\alpha_1\beta_2\beta_3\ket{011}\\
&\quad-\beta_1\alpha_2\alpha_3\ket{100}+\beta_1\alpha_2\beta_3\ket{101}
+\beta_1\beta_2\alpha_3\ket{110}-\beta_1\beta_2\beta_3\ket{111}.
\end{aligned}
\]

\paragraph{2) Expansion of \(U_1\ket{1}\otimes U_2\ket{1}\otimes U_3\ket{1}\).}
\[
\begin{aligned}
U_1\ket{1}\otimes U_2\ket{1}\otimes U_3\ket{1}
&=(\beta_1\ket{0}-\alpha_1\ket{1})\otimes(\beta_2\ket{0}-\alpha_2\ket{1})\otimes(\beta_3\ket{0}-\alpha_3\ket{1})\\
&=\beta_1\beta_2\beta_3\ket{000}-\beta_1\beta_2\alpha_3\ket{001}-\beta_1\alpha_2\beta_3\ket{010}
+\beta_1\alpha_2\alpha_3\ket{011}\\
&\quad-\alpha_1\beta_2\beta_3\ket{100}+\alpha_1\beta_2\alpha_3\ket{101}
+\alpha_1\alpha_2\beta_3\ket{110}-\alpha_1\alpha_2\alpha_3\ket{111}.
\end{aligned}
\]

\medskip
After calculating the tensor products for the probabilities of both \(\ket{0}\) and \(\ket{1}\), we disentangled the rounds to calculate the probabilities of each outcome. To disentangle the rounds, we use \(J^\dagger\), where the final amplitude \((a)\) is represented in the formulas below.

\smallskip
\textit{\(a=\) final amplitude after disentanglement, \(A=\) amplitude during the disentanglement.}

\[
\begin{aligned}
a^{\text{out}}_{000} &= \frac{1}{\sqrt{2}}\!\left(A_{000}+A_{111}\right), &\qquad
a^{\text{out}}_{111} &= \frac{1}{\sqrt{2}}\!\left(A_{111}-A_{000}\right),\\
a^{\text{out}}_{001} &= \frac{1}{\sqrt{2}}\!\left(A_{001}+A_{110}\right), &\qquad
a^{\text{out}}_{110} &= \frac{1}{\sqrt{2}}\!\left(A_{110}-A_{001}\right),\\
a^{\text{out}}_{010} &= \frac{1}{\sqrt{2}}\!\left(A_{010}+A_{101}\right), &\qquad
a^{\text{out}}_{101} &= \frac{1}{\sqrt{2}}\!\left(A_{101}-A_{010}\right),\\
a^{\text{out}}_{011} &= \frac{1}{\sqrt{2}}\!\left(A_{011}+A_{100}\right), &\qquad
a^{\text{out}}_{100} &= \frac{1}{\sqrt{2}}\!\left(A_{100}-A_{011}\right).
\end{aligned}
\]

\noindent\textit{Eg Amplitude of \(A_{000}\):} \quad
\(A_{000}=(\alpha_1\alpha_2\alpha_3+\beta_1\beta_2\beta_3)\ket{000}\).
\smallskip

\noindent Substituting yields:
\[
\begin{aligned}
\ket{\psi_{\text{unentangled}}}
&=(\alpha_1\alpha_2\alpha_3+i\beta_1\beta_2\beta_3)\ket{000}\\
&\quad+(\alpha_1\beta_2\beta_3+i\beta_1\alpha_2\alpha_3)\ket{011}\\
&\quad+(\beta_1\alpha_2\beta_3+i\alpha_1\beta_2\alpha_3)\ket{101}\\
&\quad+(\beta_1\beta_2\alpha_3+i\alpha_1\alpha_2\beta_3)\ket{110}.
\end{aligned}
\]
As you can see, 4 key notations are cancelled out.

We can interpret the wave notation this way: for any ket notation in the form \(\ket{abc}\), \(a\) is the probability of defecting at round 1, \(b\) is the probability of defecting at round 2, and \(c\) is the probability of defecting at round 3. For example \(\ket{011}\) means cooperation in round 1 and defection in rounds 2 and 3. We can utilize this fact in the following way.

We have 4 different total probabilities in our 3-round centipede game, \(P_{1d_1}\), \(P_{2d_2}\), \(P_{3d_1}\), \(P_{3c_1}\); each of these 4 probabilities can be found using the wave function formula we have found above. For example, \(P_{1d_1}\) means defection in round 1. So, we would add up the probabilities (recall that \(P(a)=\lvert a\rvert^{2}\), \(a=\) amplitude) of all of the qubit equivalents: 100, 110, 101, 111. From our final formula, we see that the coefficients of 100 and 111 are 0; thus, we only add the probabilities of 110 and 101, yielding:
\[
(\beta_1\alpha_2\beta_3)^{2}+(\alpha_1\beta_2\alpha_3)^{2}+(\beta_1\beta_2\alpha_3)^{2}+(\alpha_1\alpha_2\beta_3)^{2}.
\]
Using the same logic, we get:
\[
\begin{aligned}
P_{1d_1} &= (\beta_1\alpha_2\beta_3)^{2}+(\alpha_1\beta_2\alpha_3)^{2}+(\beta_1\beta_2\alpha_3)^{2}+(\alpha_1\alpha_2\beta_3)^{2},\\
P_{2d_2} &= (\alpha_1\beta_2\beta_3)^{2}+(\beta_1\alpha_2\alpha_3)^{2},\\
P_{3d_1} &= 0,\\
P_{3c_1} &= (\alpha_1\alpha_2\alpha_3)^{2}+(\beta_1\beta_2\beta_3)^{2}.
\end{aligned}
\]

To find the maximum payoff for each player, an equation was made by multiplying the reward of the move by the probability of getting that reward, which was represented by the decision (cooperate or defect), the player, and the round number. The maximum payoff equation of both player 1 and player 2 is shown below:

\noindent\textbf{Player 1:}
\[
\$_{1} = 1P_{1d_1} + 0P_{2d_2} + 3P_{3d_1} + 2P_{3c_1}.
\]

\noindent\textbf{Player 2:}
\[
\$_{2} = 0P_{1d_1} + 2P_{2d_2} + 1P_{3d_1} + 2P_{3c_1}.
\]
The dollar sign represents the maximum payoff for each player. The coefficient of each variable shows the payoff received by the respective player. The subscript after \(P\) shows the round number, and the \(c\)’s in rounds one and two weren’t included during the calculations because there isn’t a payoff for cooperating. Recall that we found the probability of defecting in the 3rd round was 0. Therefore, we can plug \(0\) in for \(P_{3d_1}\), and the simplified equations for the total payoff are shown below:

\noindent\textbf{Player 1:}
\[
\$_{1} = 1P_{1d_1} + (0)P_{2d_2} + 3(0) + 2P_{3c_1} = P_{1d_1} + 2P_{3c_1}.
\]

\noindent\textbf{Player 2:}
\[
\$_{2} = 0P_{1d_1} + 2P_{2d_2} + 1(0) + 2P_{3c_1} = 2P_{2d_2} + 2P_{3c_1}.
\]

To find the optimal average payoff, we need to find the partial derivative of the two maximum payoff equations with respect to the three \(\theta\) values.
\[
\frac{\partial \$_{1}}{\partial \theta_{1}}=0,\quad
\frac{\partial \$_{1}}{\partial \theta_{2}}=0,\quad
\frac{\partial \$_{1}}{\partial \theta_{3}}=0,\quad
\frac{\partial \$_{2}}{\partial \theta_{1}}=0,\quad
\frac{\partial \$_{2}}{\partial \theta_{2}}=0,\quad
\frac{\partial \$_{2}}{\partial \theta_{3}}=0.
\]

Our approach is to find the partial derivative of each probability with respect to the three different theta values. By factoring and employing cosine double-angle formulas and Pythagorean identities, we yield:
\[
\frac{\partial \$_{1}}{\partial \theta_{1}}
= 2\cos\!\left(\frac{\theta_{1}}{2}\right)\sin\!\left(\frac{\theta_{1}}{2}\right)
\Bigl(-\cos^{2}\!\left(\frac{\theta_{2}}{2}\right)\cos^{2}\!\left(\frac{\theta_{3}}{2}\right)
+ \sin^{2}\!\left(\frac{\theta_{2}}{2}\right)\sin^{2}\!\left(\frac{\theta_{3}}{2}\right)\Bigr)=0,
\]
\[
\frac{\partial \$_{1}}{\partial \theta_{2}}
= -\cos\!\left(\frac{\theta_{2}}{2}\right)\sin\!\left(\frac{\theta_{2}}{2}\right)\cos(\theta_{1})=0,
\]
\[
\begin{aligned}
\frac{\partial \$_{1}}{\partial \theta_{3}}
&= \cos\!\left(\frac{\theta_{3}}{2}\right)\sin\!\left(\frac{\theta_{3}}{2}\right)
\Bigl(\cos^{2}\!\left(\frac{\theta_{2}}{2}\right)-\sin^{2}\!\left(\frac{\theta_{2}}{2}\right)\\
&\qquad\quad
-2\cos^{2}\!\left(\frac{\theta_{1}}{2}\right)\cos^{2}\!\left(\frac{\theta_{2}}{2}\right)
+2\sin^{2}\!\left(\frac{\theta_{1}}{2}\right)\sin^{2}\!\left(\frac{\theta_{2}}{2}\right)\Bigr)=0.
\end{aligned}
\]

\[
\frac{\partial \$_{2}}{\partial \theta_{1}}
= \cos\!\left(\frac{\theta_{1}}{2}\right)\sin\!\left(\frac{\theta_{1}}{2}\right)\cdot 0
=0,\ \text{therefore }\theta_{1}\text{ can be any value.}
\]
\[
\frac{\partial \$_{2}}{\partial \theta_{2}}
= -2\cos\!\left(\frac{\theta_{2}}{2}\right)\sin\!\left(\frac{\theta_{2}}{2}\right)\cos(\theta_{3})=0,
\qquad
\frac{\partial \$_{2}}{\partial \theta_{3}}
= -2\cos\!\left(\frac{\theta_{3}}{2}\right)\sin\!\left(\frac{\theta_{3}}{2}\right)\cos(\theta_{2})=0.
\]

For \(\dfrac{\partial \$_{1}}{\partial \theta_{1}}\) and \(\dfrac{\partial \$_{1}}{\partial \theta_{3}}\) some parts can't be simplified further, although the theta values are the same. We can put the two formulas into a system of equations. Since there are only two formulas with three variables, we will not be able to solve for the variables.

\paragraph{The resulting theta values are:}
\[
\theta_{1}\in\left\{0,\ \pi,\ \frac{\pi}{2}\right\} \ \text{(taken from }\$_{1},\ \text{since only player 1 can decide for round 1 (\(\theta_{1}\)))} 
\]
\[
\theta_{2}\in\left\{0,\ \pi,\ \frac{\pi}{2}\right\} \ \text{(taken from }\$_{2},\ \text{for the same reason)}
\]
\[
\theta_{3}\in\left\{0,\ \pi\right\} \ \text{(taken from }\$_{1},\ \text{for the same reason)}.
\]

\medskip
\noindent\textit{To save time and our sanity, instead of substituting each answer, we used simulation with Qiskit to find which combination would give us the optimal output.}

\subsection*{Simulation}
To simulate our algorithm, we used Google Colab and coded in Python with a quantum framework called Qiskit. Qiskit is an open-source software development kit, primarily built by IBM [21]. It allows users to build, run, and optimize simple quantum circuits and algorithms on real quantum hardware and simulations.

Using Qiskit, we simulated our adopted algorithm using a three-qubit quantum circuit. Each qubit represented one round within the centipede game. Players’ strategies were encoded using \(R_y(-\theta)\) rotations, with \(\theta \in \{0, \pi/2, \pi\}\), and the decisions were made (rotations were applied) while the three qubits were in a state of entanglement. Although the centipede game is played sequentially, our game model in Python applies the quantum strategies all at once, as seen in Figure 2. This is because quantum mechanics requires the entire system to evolve simultaneously and in a unified manner. Each round's strategy influences the quantum system as a whole, rather than being based on the outcomes of previous rounds, as would be the case in classical turn-based games. As discussed in previous works, such as [22], quantum game models require that strategies be applied all at once due to the inherent nature of quantum evolution and measurement. Even if we had applied the decisions one at a time, the resulting payoffs would have reflected the same results. After the strategies were applied, each qubit was disentangled and then measured to find the outcome of the game. 

\medskip
\begin{figure}[H]
    \centering
    \includegraphics[width=0.5\linewidth]{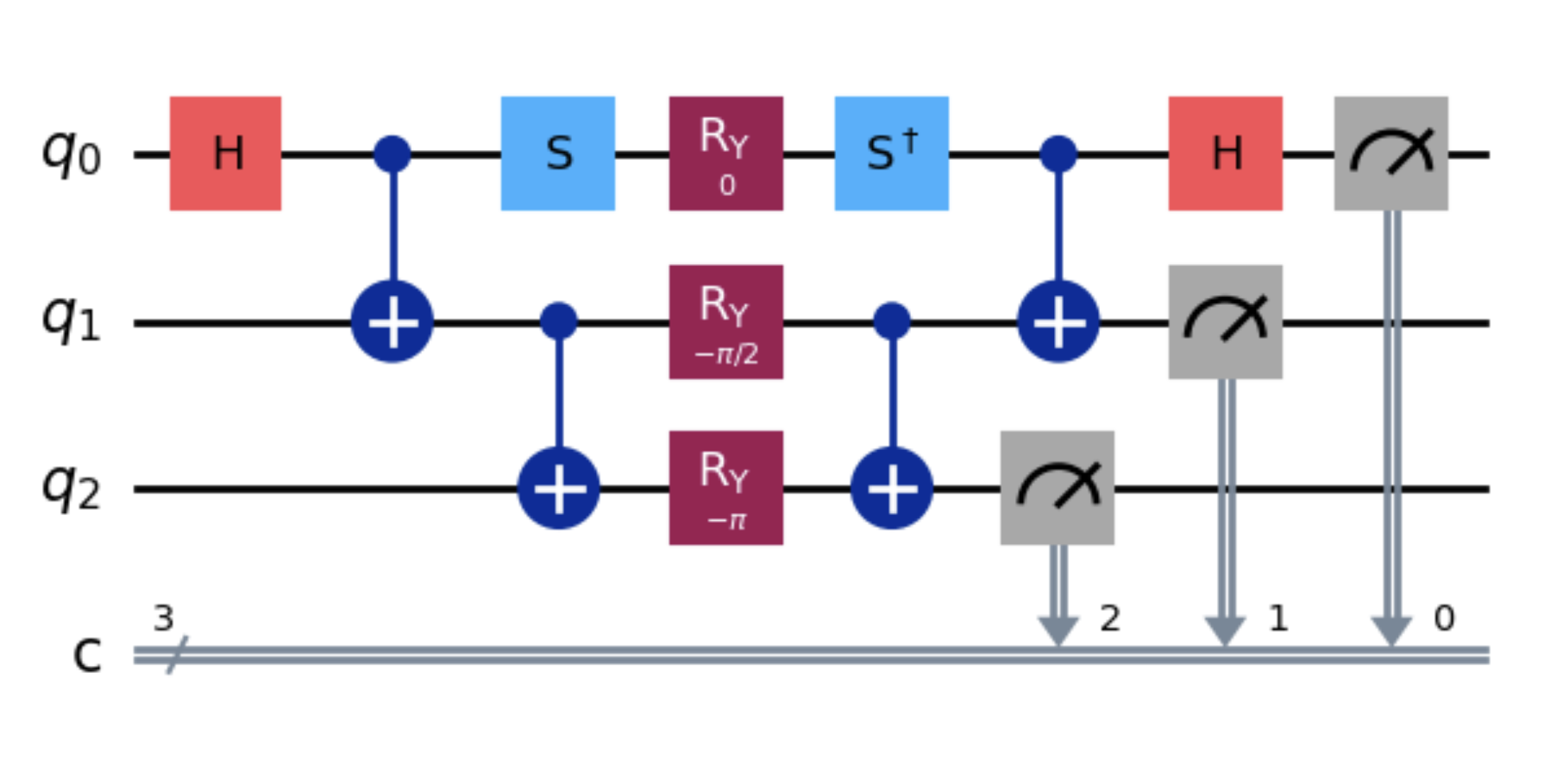}
    \caption{The quantum circuit above is the strategy circuit of a 3-round quantum centipede game. The \(q_0, q_1, q_2\) represent the 3 rounds of the centipede game. The \(H\) and CNOT gates create entanglement, the \(S\) gate adds the necessary phase difference, \(R_y\) boxes apply the \(y\) rotation, and the \(c\) line stores the results of the quantum measurements. Qiskit was used to visualize the circuit.}
    \label{fig:qcentipede}
\end{figure}
\medskip

Using the equation derived in the previous section, we calculated the expected payoff for each round. For each combination of \(\theta\) values, we ran 1000 trials of the game and found the average payoff for each strategy applied. The resulting payoffs are shown in the graphs below.

\section{RESULTS}
We used the Eisert–Wilkens–Lewenstein protocol to create a quantum algorithm in order to find a more mutually beneficial Nash equilibrium for the centipede game. We mathematically calculated the possible \(\theta\) values that could serve as player strategies and derived the equations for each player’s payoff. We then simulated 1000 games for each combination of \(\theta\) values (strategies) of the centipede game using a Python library called Qiskit and recorded the average payoff for each strategy. Table 1 shows the final simulation result and the total payoff for each.

\begin{table}[ht]
\centering
\caption{Raw averages of the payoffs for each strategy}
\label{tab:raw-averages}
\begin{tabular}{>{\centering\arraybackslash}m{0.48\linewidth} >{\centering\arraybackslash}m{0.32\linewidth}}
\toprule
\textbf{Strategy \([\theta_1,\theta_2,\theta_3]\)} & \textbf{Payoff \([\$_1,\$_2]\)} \\
\midrule
\fbox{\([0,0,0]\)}                           & \fbox{\([2.0,\,2.0]\)} \\
\([\pi,0,0]\)                                 & \([0.0,\,2.0]\) \\
\([\pi/2,0,0]\)                               & \([0.96,\,2.0]\) \\
\([0,\pi,0]\)                                 & \([1.0,\,0.0]\) \\
\([\pi,\pi,0]\)                               & \([1.0,\,0.0]\) \\
\([\pi/2,\pi,0]\)                             & \([1.0,\,0.0]\) \\
\([0,\pi/2,0]\)                               & \([1.51,\,1.02]\) \\
\([\pi,\pi/2,0]\)                             & \([0.54,\,0.92]\) \\
\([\pi/2,\pi/2,0]\)                           & \([0.92,\,0.96]\) \\
\([0,0,\pi]\)                                 & \([1.0,\,0.0]\) \\
\([\pi,0,\pi]\)                               & \([1.0,\,0.0]\) \\
\([\pi/2,0,\pi]\)                             & \([1.0,\,0.0]\) \\
\([0,\pi,\pi]\)                               & \([0.0,\,2.0]\) \\
\fbox{\([\pi,\pi,\pi]\)}                      & \fbox{\([2.0,\,2.0]\)} \\
\([\pi/2,\pi,\pi]\)                           & \([1.08,\,2.0]\) \\
\([0,\pi/2,\pi]\)                             & \([0.44,\,1.12]\) \\
\([\pi,\pi/2,\pi]\)                           & \([1.56,\,1.12]\) \\
\([\pi/2,\pi/2,\pi]\)                         & \([1.05,\,1.14]\) \\
\bottomrule
\end{tabular}
\end{table}

The \(\theta\) values \([0, 0, 0]\) and \([\pi, \pi ,\pi]\) were found to have the most payoff per player, demonstrating two new Nash equilibria. These are the global maxima that we found after taking the partial derivatives of the average payoff equations. The other inputs we tried were critical points resulting from partial differentiation. However, they are extraneous solutions because the average payoff is below \(2\) for either one or both of the players. 
As discussed above in the math section, we thought that only the combination of \(\theta\) values \((0, 0, 0)\) would be the solution with the best payoff for both players, because the values indicate (classically) that players would cooperate in all three rounds. 
Our findings contradict our original assumptions. Our simulation indicated that both \((0,0,0)\) and \((\pi,\pi,\pi)\) yielded the best outcome, which is to cooperate. How can there be \(2\) solutions when there is only \(1\) classical way to produce a payoff of \((2,2)\)? What is going on here? 

To verify our result, we can substitute the \(\theta\) values into the 
\((\pi,\pi,\pi)\), where \(\theta_1=\theta_2=\theta_3=\pi \Rightarrow \alpha_1=\alpha_2=\alpha_3=0,\ \beta_1=\beta_2=\beta_3=1\)
\[
\lvert\psi_{\text{unentangled}}\rangle=(0+i\cdot 1)\lvert 000\rangle+(0+i\cdot 0)\lvert 011\rangle+(0+i\cdot 0)\lvert 101\rangle+(0+i\cdot 0)\lvert 110\rangle=i\,\lvert 000\rangle
\]
Similarly
\((0,0,0)\), where \(\theta_1=\theta_2=\theta_3=0\)
\[
\lvert\psi_{\text{unentangled}}\rangle=(1+i\cdot 0)\lvert 000\rangle=\lvert 000\rangle
\]
Although the player strategies differ, both sets of angles produce the same measurable state \(\lvert 000\rangle\), corresponding to “cooperate–cooperate–cooperate.” The reason lies in the structure of the entangled wavefunction. The interference terms in the GHZ-type state cause the amplitudes associated with “defect in the final round” to cancel exactly, giving \(P_{3d1}=0\). Once \(P_{3d1}=0\), the branch of the wavefunction corresponding to last-round defection disappears entirely, leaving only the cooperative branch \(\lvert 000\rangle\), successfully avoiding the classical backwards induction trap.
By the wizardry of quantum mechanics, this leads to a remarkable symmetry in the strategy space. If we imagine all possible strategies as points in a three-dimensional cube with sides of length \(\pi\), the two opposite corners, \((0,0,0)\) and \((\pi,\pi,\pi)\), are mirror images of one another. They differ only by a relative phase factor \(i\), which is physically unobservable. This phase symmetry makes both corners equivalent in terms of measurable outcomes, a fact confirmed by our simulation results. In summary, the player strategies can affect each other in the quantum centipede game in ways that are classically not possible, demonstrating that the quantum centipede game produces new cooperative Nash equilibrium.


\section{DISCUSSION}
Unlike classical game theory, which relies on fixed strategies and probability distributions, quantum game theory introduces a broader framework by incorporating principles of quantum mechanics like superposition and entanglement. Quantizing the centipede game allows for sustained cooperation. Entanglement correlates decisions across rounds, effectively binding the players’ strategies together. Unlike classical reasoning, where each round is an isolated choice, our model creates a unified strategic state, preventing the backward induction logic that drives immediate defection. 

To our knowledge, we present the first application of the continuous Eisert–Wilkens–Lewenstein protocol with maximal entanglement and full strategy space to a multi-round game, as well as the first practical implementation of a quantum centipede game on Qiskit, an IBM backend quantum simulation, to confirm our findings. We expanded the set of possible outcomes and player strategies, allowing entirely new approaches and types of equilibria that go beyond classical limits. Our simulation revealed two coherent quantum Nash equilibrium profiles, both of which represent the classical path of cooperate-cooperate-cooperate. This outcome demonstrates a degeneracy: although the wave function for \((\pi, \pi, \pi)\) differs from \((0, 0, 0)\), both strategies' payoffs are equivalent. 

Based on the findings of our research, we have constructed the following conjecture.
\subsection*{Conjecture}

\begin{description}[leftmargin=0pt,labelsep=0.75em]
\item[\textbf{Preconditions.}]
In a 2-player, sequential, 2-choice game (cooperate/defect) with \(n\ge 2\) rounds, index rounds so that the terminal move is round \(n\).
Assume:
\begin{enumerate}\itemsep4pt
    \item[(i)] The classical SPNE is obtained by backward induction.
    \item[(ii)] The game is quantized in the EWL/CTZ style: each round is a qubit \((\lvert 0\rangle=\text{cooperate},\ \lvert 1\rangle=\text{defect})\); players apply local unitaries; rounds are entangled; measurement is in the computational basis.
    \item[(iii)] Payoffs are assigned to the resulting bit string in the usual way.
\end{enumerate}

\textbf{If these preconditions are satisfied, then}
\begin{enumerate}\itemsep8pt
    \item \textbf{Backward induction collapse.} The probability of defecting at the last round \(m\) is zero, breaking the classical backward-induction dilemma:
    \[
      P_{\text{defect}}(m)=0.
    \]
    \item \textbf{Degenerate cooperative equilibria.} Entanglement can produce symmetry in outcomes, allowing multiple classical input configurations, e.g., \((0,0,0)\) and \((\pi,\pi,\pi)\), to yield practically identical quantum states \(i\lvert 000\rangle\) and \(\lvert 000\rangle\), and thus identical cooperative payoffs. Hence the quantum game admits corner-degenerate equilibria in which distinct parameter sets correspond to equivalent cooperative outcomes.
\end{enumerate}
\end{description}

We are not completely sure why the symmetry exists, nor what specific type of symmetry it is (from coordinate \((0, 0, 0)\) to \((\pi, \pi, \pi)\), is it a geometric reflection in strategy space?). This is where we acknowledge our limited knowledge as high school students in this field and strongly encourage researchers like those of you reading right now to tackle this conjecture and contribute to the field of quantum game theory. 

Lastly, our research also demonstrates how certain qubits may carry only phase information that does not affect the final measurement. In our case, since \(P_{3d1}=0\), the last qubit is always measured as \(\lvert 0\rangle\) (cooperation). Its associated unitary operator can therefore be taken as the identity \((U_3=I)\), meaning the last qubit no longer influences the calculation or payoff structure.

These strategies would not exist in a classical framework, yet they mirror choices frequently observed in experiments. The centipede game, in particular, has long demonstrated a disconnect between theoretical predictions and observed human behavior. Classical theory predicts immediate defection, while experimental studies show sustained cooperation. Our study shows how quantum algorithms can better model how the game plays out in real life. According to a study done by Paul J. Healy [23], no players chose to defect in the first three rounds, while around 31 percent of players chose to split the pot at the end of the game. Another study [24] shows that in the early rounds of the centipede game, cooperation rates were around 70-75 percent, reinforcing the idea that classical models alone fail to accurately capture human behavior. Our findings suggest that quantum game theory is a promising tool for bridging this gap.
These contributions validate the feasibility of implementing quantum game protocols practically. They also open new pathways to applying quantum games to models of real-life interactions. Game theory has many different applications in many fields. Historically, game theory has been used by economists to study the effects of self-interest, rationality, and equilibrium in situations where traditional price theory does not apply. It models strategic interactions between different corporations, clients, and consumers by using simple games, such as those we studied in this paper [25]. Beyond economics, game theory is also used to model legal decisions and explain how people make choices that affect both themselves and others. However, these classical models often struggle to accommodate decisions that don’t align closely with achieving maximal payoff. By introducing quantum strategies to solve these smaller games and create non-classical forms of equilibria, our work offers a better way to capture the nuances of human behavior and may serve as a contributor to next-generation tools in behavioral science, economics, and beyond. 
Our study is limited to a 3-round centipede game due to computational limits and complexity. A possible extension of this paper could be simulating multiple rounds and seeing if our hypotheses would still hold, or applying our protocol to more complex game theory games. We also simulated our algorithm on an ideal, noiseless backend. Noise on a suboptimal simulation could be introduced to see how it would affect the calculations of the algorithm. This goes hand-in-hand with simulating our algorithm on real quantum hardware once the technology supports it. Real quantum computers would be affected by their surroundings, and things like noise, errors, and unstable qubits could change the results of the game. Testing our algorithm on real hardware would help us see if it still works well outside of a perfect simulation.


\sloppy
\section{REFERENCES}
\begin{enumerate}[label={[}\arabic*{]}]
\item J. v. Neumann, “Zur Theorie der Gesellschaftsspiele, Mathematische Annalen", vol. 100, No. 1, pp. 295–320, Dec. 1928, doi: \href{https://doi.org/10.1007/bf01448847}{https://doi.org/10.1007/bf01448847}. [Accessed: Jul. 12, 2025].
\item T. Roughgarden, “CS261: A Second Course in Algorithms Lecture \#10: The Minimax Theorem and Algorithms for Linear Programming * 1 Zero-Sum Games and the Minimax Theorem 1.1 Rock-Paper Scissors,” 2016. Available: \url{https://theory.stanford.edu/~tim/w16/l/l10.pdf}. [Accessed: Jul. 20, 2025].
\item R. Sekhon and R. Bloom, “4.2: Maximization By The Simplex Method,” Mathematics LibreTexts, Mar. 22, 2020. \url{https://math.libretexts.org/Bookshelves/Applied_Mathematics/Applied_Finite_Mathematics_(Sekhon_and_Bloom)/04\%3A_Linear_Programming_The_Simplex_Method/4.02\%3A_Maximization_By_The_Simplex_Method}. [Accessed: Jul. 20, 2025].
\item R. A. Forder, “Military Operations Research: Quantitative Decision Making,” Journal of the Operational Research Society, vol. 49, no. 11, pp. 1227–1228, Nov. 1998, doi: \href{https://doi.org/10.1057/palgrave.jors.2600040}{https://doi.org/10.1057/palgrave.jors.2600040}. [Accessed: Jul. 12, 2025].
\item C. E. Lemke and J. T. Howson, “Equilibrium Points of Bimatrix Games,” Journal of the Society for Industrial and Applied Mathematics, vol. 12, no. 2 (Jun., 1964), pp. 413-423, Mar. 06, 2005. \url{http://links.jstor.org/sici?sici=0368-4245%28196406%2912%3A2%3C413%3AEPOBG%3E2.0.CO%3B2-R}. [Accessed: Jul. 13, 2025].
\item C. A. Holt and A. E. Roth, “The Nash equilibrium: A perspective,” Proceedings of the National Academy of Sciences, vol. 101, no. 12, pp. 3999–4002, Mar. 2004, doi: \href{https://doi.org/10.1073/pnas.0308738101}{https://doi.org/10.1073/pnas.0308738101}. [Accessed: Jul. 2, 2025].
\item D. Styer, “A Brief History of Quantum Mechanics,” \url{https://www2.oberlin.edu/physics/dstyer/StrangeQM/history.html}, 1999. [Accessed: Jul. 12, 2025].
\item R. P. Feynman, “Simulating physics with computers,” International Journal of Theoretical Physics, vol. 21, no. 6–7, pp. 467–488, Jun. 1982, doi: \href{https://doi.org/10.1007/bf02650179}{https://doi.org/10.1007/bf02650179}. [Accessed: Jul. 9, 2025].
\item I. Ghosh, “Quantum Game Theory — I,” Resonance, vol. 26, no. 5, pp. 671–684, May 2021, doi: \href{https://doi.org/10.1007/s12045-021-1168-2}{https://doi.org/10.1007/s12045-021-1168-2}. [Accessed: Jul. 2, 2025].
\item D. A. Meyer, “Quantum strategies,” arXiv:quant-ph/9804010 [quant-ph], Apr. 1998. [Online]. Available: \url{https://arxiv.org/pdf/quant-ph/9804010}. [Accessed: Jul. 16, 2025].
\item S. Das, “Quantumizing Classical Games: An Introduction to Quantum Game Theory,” arXiv:2305.00368v1 [quant-ph], Apr. 2023. [Online]. Available: \url{https://arxiv.org/pdf/2305.00368v1}. [Accessed: Jul. 16, 2025].
\item S. E. Landsburg, “Quantum Game Theory,” Wiley Encyclopedia of Operations Research and Management Science, Jan. 2011, doi: \href{https://doi.org/10.1002/9780470400531.eorms0697}{https://doi.org/10.1002/9780470400531.eorms0697}. [Accessed: Jul. 11, 2025].
\item Economic Applications of Game Theory, Chapter 9: Backward Induction, MIT OpenCourseWare, Prof. Muhamet Yildiz, Fall 2012. [Online]. Available: \url{https://ocw.mit.edu/courses/14-12-economic-applications-of-game-theory-fall-2012/4b4412575dc74593c9d9c59e94427b69_MIT14_12F12_chapter9.pdf}. [Accessed: Jul. 4, 2025].
\item ELMSHAUSER Béla, “Altruism and Ambiguity in the Centipede game,” Oct. 2022, doi: \href{https://doi.org/10.31235/osf.io/93p8s}{https://doi.org/10.31235/osf.io/93p8s}. [Accessed: Jul. 10, 2025].
\item R. D. McKelvey and T. R. Palfrey, “An Experimental Study of the Centipede Game,” Econometrica, vol. 60, no. 4, p. 803, Jul. 1992, doi: \href{https://doi.org/10.2307/2951567}{https://doi.org/10.2307/2951567}. [Accessed: Jul. 2, 2025].
\item D. R. Simon, “On the Power of Quantum Computation,” in \textit{Proc. 35th Annu. Symp. on Foundations of Computer Science (FOCS)}, Santa Fe, NM, Nov. 1994, pp. 116–123. [Online]. Available: \url{https://www.researchgate.net/profile/Daniel-Simon-11/publication/2822536_On_the_Power_of_Quantum_Computation/links/55ccbf3608aeca56cc1c185/On-the-Power-of-Quantum-Computation.pdf}. [Accessed: Jul. 2, 2025].
\item R. Ibarrondo, G. Gatti, and M. Sanz, “Quantum vs classical genetic algorithms: A numerical comparison shows faster convergence,” 2021 IEEE Symposium Series on Computational Intelligence (SSCI), Dec. 2022, doi: \href{https://doi.org/10.1109/ssci51031.2022.10022159}{https://doi.org/10.1109/ssci51031.2022.10022159}. [Accessed: Jul. 2, 2025].
\item V. R. Rao, N. Yang, and S. Zakerinia, “Using Quantum Game Theory to Model Competition,” 2024, doi: \href{https://doi.org/10.2139/ssrn.5048345}{https://doi.org/10.2139/ssrn.5048345}. [Accessed: Jul. 12, 2025].
\item J. Eisert, M. Wilkens, and M. Lewenstein, “Quantum Games and Quantum Strategies,” Physical Review Letters, vol. 83, no. 15, pp. 3077–3080, Oct. 1999, doi: \href{https://doi.org/10.1103/physrevlett.83.3077}{https://doi.org/10.1103/physrevlett.83.3077}. [Accessed: Jul. 2, 2025].
\item P. Frackiewicz, “A new quantum scheme for normal-form games,” arXiv:1701.07096 [quant-ph], Jan. 2017. [Online]. Available: \url{https://arxiv.org/pdf/1701.07096}. [Accessed: Jul. 12, 2025].
\item “Introduction to Qiskit | IBM Quantum Documentation,” IBM Quantum Documentation, 2017. \url{https://quantum.cloud.ibm.com/docs/en/guides}. [Accessed: Jul. 10, 2025].
\item D. Aerts, “Towards a quantum evolutionary scheme: violating Bell’s inequalities in language,” arXiv:quant-ph/0407150 [quant-ph], Jul. 2004. [Online]. Available: \url{https://arxiv.org/pdf/quant-ph/0407150}.  [Accessed: Jul. 8, 2025].
\item P. Healy et al., “EPISTEMIC EXPERIMENTS: UTILITIES, BELIEFS, AND IRRATIONAL PLAY † and Michigan State. I have benefited greatly from conversations with.” Accessed: Jul. 16, 2025. [Online]. Available: \url{https://healy.econ.ohio-state.edu/papers/Healy-EpistemicExperiments.pdf}
\item P. Dal Bó and G. R. Fréchette, “The Evolution of Cooperation in Infinitely Repeated Games: Experimental Evidence,” American Economic Review, vol. 101, no. 1, pp. 411–429, Feb. 2011, doi: \href{https://doi.org/10.1257/aer.101.1.411}{https://doi.org/10.1257/aer.101.1.411}. [Accessed: Jul. 14, 2025].
\item R. Gibbons, “An Introduction to Applicable Game Theory,” Journal of Economic Perspectives, vol. 11, no. 1, pp. 127–149, Feb. 1997, doi: \href{https://doi.org/10.1257/jep.11.1.127}{https://doi.org/10.1257/jep.11.1.127}. [Accessed: Jul. 7, 2025].
\end{enumerate}
\fussy

\section{ACKNOWLEDGEMENTS}
We would like to thank Ariana Aghamohammadi for her insightful lectures and wisdom. We would also like to thank our Teaching Assistants, Parsa Madinei, Manan Gupta, and Sina Ahadi, for providing us with guidance and advice on our work. We acknowledge the Summer Research Academies program, as well as Dr. Lina Kim, for providing us with the opportunity to conduct professional research. 

\section{AUTHOR CONTRIBUTION STATEMENT}
All authors conceived the research question, designed the algorithm, and performed the computations, simulations, and analysis. All authors contributed to the first draft of the manuscript. For validation, Yiguo Zhang reproduced the derivations, resolved mathematical issues, and updated the methodological framework. Kaytki Chakankar and Yiguo Zhang then managed revisions and finalized the manuscript. Grammarly was used to ensure grammatical consistency.

\end{document}